\documentclass[english,prl,twocolumn,showpacs,amsmath,amssymb,superscriptaddress,floatfix]{revtex4}

\usepackage[T1]{fontenc}
\usepackage{graphicx}
\usepackage[latin9]{inputenc}
\usepackage{amsmath}
\usepackage{amssymb}
\usepackage{esint}

\makeatletter
\gdef\@ptsize{0}
\let\@cursize\normalize
\makeatother
\usepackage{setspace}

\usepackage{bm}
\usepackage{pifont}
\usepackage{babel}
\makeatother

\begin{document}

\preprint{}

\title{Single Spin Detection with a Carbon Nanotube Double Quantum Dot}

\author{S. J. Chorley}
\affiliation{Cavendish Laboratory, University of Cambridge, JJ Thomson Avenue,
Cambridge, CB3 0HE, United Kingdom}

\author{G. Giavaras}
\author{J. Wabnig}
\affiliation{Department of Materials, University of Oxford, Parks Road, Oxford,
OX1 3PH, United Kingdom}

\author{G. A. C. Jones}
\author{C. G. Smith}
\affiliation{Cavendish Laboratory, University of Cambridge, JJ Thomson Avenue,
Cambridge, CB3 0HE, United Kingdom}

\author{G. A. D. Briggs}
\affiliation{Department of Materials, University of Oxford, Parks Road, Oxford,
OX1 3PH, United Kingdom}

\author{M. R. Buitelaar}
\email{mrb51@cam.ac.uk}
\affiliation{Cavendish Laboratory, University of Cambridge, JJ Thomson Avenue,
Cambridge, CB3 0HE, United Kingdom}

\date{\today}


\maketitle


\textbf{Spin qubits defined in carbon nanotube quantum dots are of considerable interest for encoding and manipulating quantum information because of the long electron spin coherence times expected. However, before carbon nanotubes can find applications in quantum information processing schemes, we need to understand and control the coupling between individual electron spins and the interaction between the electron spins and their environment. Here we make use of spin selection rules to directly measure - and demonstrate control of - the singlet-triplet exchange coupling between two carbon nanotube quantum dots. We furthermore elucidate the effects of spin-orbit interaction on the electron transitions and investigate the interaction of the quantum dot system with a single impurity spin - the ultimate limit in spin sensitivity.}

\begin{figure*}
\includegraphics[width=150mm]{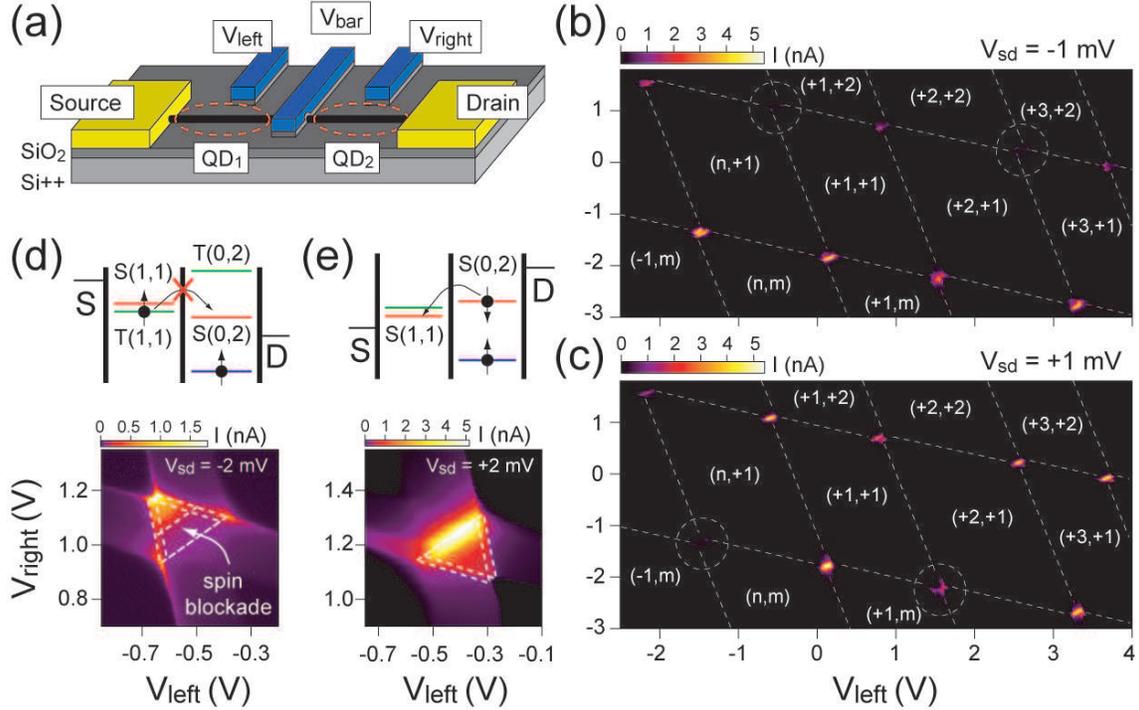}
\caption{\label{Fig1}\textbf{(a)} Schematic of the
carbon nanotube double quantum dot device used in the experiment. \textbf{(b)} Charge stability diagram of the double
dot for $V_{\textrm{sd}}=-1$ mV. The circles indicate the regions
of spin blockade. \textbf{(c)} Charge stability diagram of the same
region for $V_{\textrm{sd}}=+1$ mV. \textbf{(d,e)} Detailed measurements
of the bias triangles at the (0,1) to (1,2) charge transition for both polarities. Note the applied bias of 2 mV which allows us to observe the first excited states in the bias triangles, and the difference in current scales for the two figures. Since the inter-dot capacitance is small, the bias triangles for the electron and hole cycle strongly overlap. The energy level diagrams illustrate possible and forbidden transitions in the double quantum dot.}
\end{figure*}

A powerful method to probe the spin dynamics of quantum dots is by measuring electron transport in a double quantum dot device in the spin blockade regime \cite{Ono}. In this transport regime, the tunneling of an electron between the two quantum dots is forbidden by spin selection rules and hence the current is suppressed. However, spin blockade can be lifted by the interaction of the electron spins with their environment and a measurement of the (leakage) current thus directly probes these interactions \cite{Koppens, Johnson, Petta}. The main spin relaxation and decoherence modes in carbon nanotube that have been considered so far are hyperfine and spin-orbit coupling. The importance of hyperfine interaction in the electron spin dynamics of carbon nanotubes has recently been demonstrated in $^{13}$C enriched nanotube quantum dots \cite{Churchill1,Churchill2}. The significance of spin-orbit coupling, a result of the curvature of the carbon sheet, was demonstrated for carbon nanotubes \cite{Kuemmeth, Ando, Huertas, Bulaev} and is expected to be relevant for graphene as well \cite{Huertas}.

A further important consideration in any realistic device is the presence of impurities or defects and their coupling to the electron spins. For example, magnetic catalyst particles are used in nearly all carbon nanotube synthesis methods while defects such as vacancies are also widely regarded to give rise to magnetic behavior in carbon materials \cite{Ma, Peres, Yazyev}. While defect densities in high-quality carbon nanotubes are extremely low, their presence is still highly relevant in the context of quantum information processing in which the coupling to even a single impurity spin will affect device performance. A single impurity spin, however, is extremely difficult to detect. Here we achieve this by using the exceptional sensitivity of the transport current of a double quantum dot in the spin blockade regime to the spin environment. We show that the interplay of an impurity spin and spin-orbit interaction has a dramatic effect on the spin states of the double dot and find excellent agreement with a theoretical model.

The device we consider is a single-walled carbon nanotube grown by
chemical vapor deposition using methane with natural isotope ratios
and contacted by Au electrodes. Side and top barrier
gates are used to define and control the double quantum dot, see Fig.
1(a). A typical charge stability diagram of the device is
shown in Figs. 1(b,c) in which the ordered pairs ($n,m$) indicate
the effective electron occupancies of the many-electron double quantum
dot. In the presence of a source-drain bias voltage of $V_{\textrm{sd}}=-1$
mV, a honeycomb structure, characteristic of a double quantum dot
\cite{Mason, Sapmaz,Graber} is clearly visible. The large-small-large-small alternation
of the addition energy in the honeycomb pattern indicates that the
electron states are spin degenerate but that the orbital degeneracy
of the nanotube \cite{Buitelaar, Liang, Cobden} has been broken.

Of particular interest is the observation of Pauli spin blockade \cite{Buitelaar2},
of which a characteristic feature is the strong bias dependence of
the current for every other added electron as seen in the top and
bottom rows of Figs. 1(b,c), respectively. This is further illustrated by the detailed
measurements in Figs. 1(d,e) that correspond to the region in Figs.
1(b,c), highlighted by the dashed circle, in which the excess number
of electrons changes from (0,1) to (1,2). For negative bias voltage
the current is strongly suppressed while a large current is observed
at the base of the bias triangles when the bias is positive.

These measurements can be understood considering that, for negative bias, a flow of
electrons from the left to the right quantum dot necessarily
involves a transition from the (1,1) to the (0,2) charge state. Since
the (0,2) ground state has to be a singlet by virtue of the Pauli
exclusion principle, the (1,1)$\rightarrow$(0,2) transition is forbidden
by spin selection rules when the electrons on the double dot form
a T(1,1) triplet state, see also schematics in Figs. 1(d,e). When
the bias is positive the S(0,2) $\rightarrow$ S(1,1) transition
involves singlet states only and electrons can freely move from the
right to the left quantum dot and no current suppression is observed.

\begin{figure*}
\includegraphics[width=165mm]{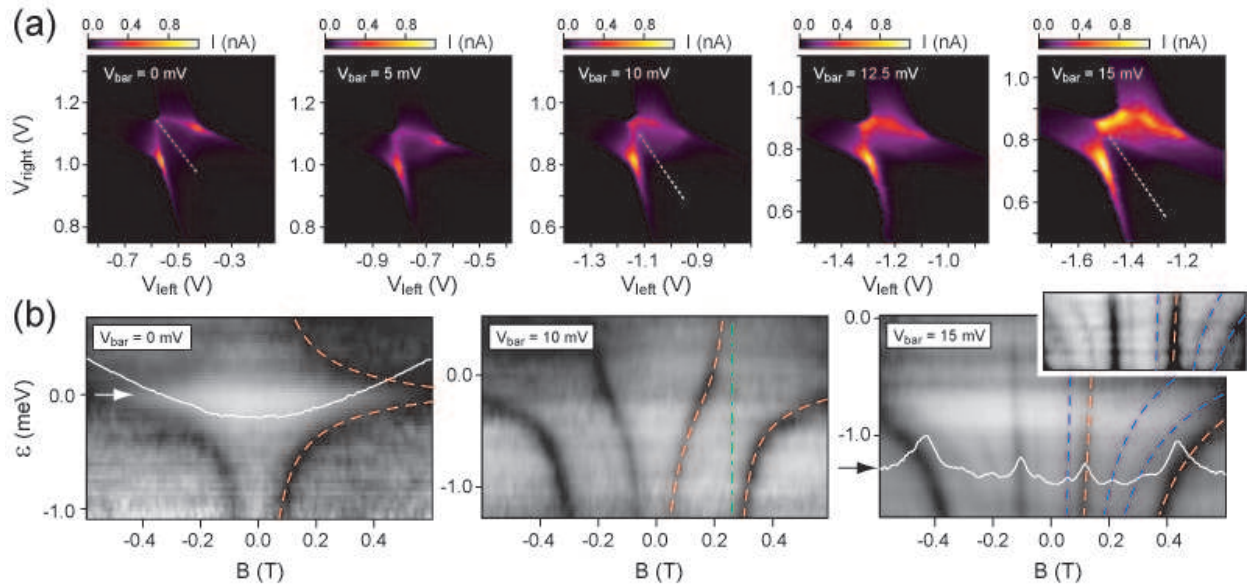}
\caption{\label{Fig2}\textbf{(a)} Bias triangles at the (0,1) to (1,2) charge
transition for $V_{\textrm{sd}}=-1$ mV and $B$ = 0 T for five
different barrier gate voltages. \textbf{(b)} Normalized current as
a function of detuning and magnetic field for $V_{\textrm{bar}}$
= 0, -10, and -15 mV. The detuning (energy difference between the
(1,1) and (0,2) charge states) axes follow the dashed white lines in panel
(a) for the three respective barrier gate voltages. In the leftmost
panel two sets of curves of high current are visible. In the middle
panel an additional set of curves appears close to zero magnetic field.
The rightmost panel shows a series of additional faintly visible current
peaks as illustrated by the linetrace and the inset which shows part of the measurement ($|B| \leq 0.37$ T) with enhanced contrast. The evolution with detuning
and magnetic field of the features closely follows that
predicted by the model described in the main text as illustrated by the dashed curves. The free fitting parameters are the interdot tunnel coupling and the strength of the Heisenberg interaction between the double dot and a spin-1/2.}
\end{figure*}

In the following we make use of spin selection rules to directly probe the spin system of the
double quantum dot in detail. Our main results are illustrated in Fig.~2.
The top row shows the stability diagrams of spin blockaded bias triangles
for five different barrier potentials. The bottom row shows the current
through the double dot as a function of magnetic field $B$ and detuning $\epsilon$
for three fixed barrier potentials, $V_{\textrm{bar}}=$0,10, and
15 mV, corresponding to three of the stability diagrams in Fig.~2(a).
Note that in Fig.~2(b) we plot the \textit{normalized}
current to accentuate the evolution of the spin states by effectively
subtracting the background current. The complete measurement set,
for all measured barrier voltages, as well as examples of measurements
at other bias triangles are shown in the Supplementary Information.

\begin{figure*}
\includegraphics[width=165mm]{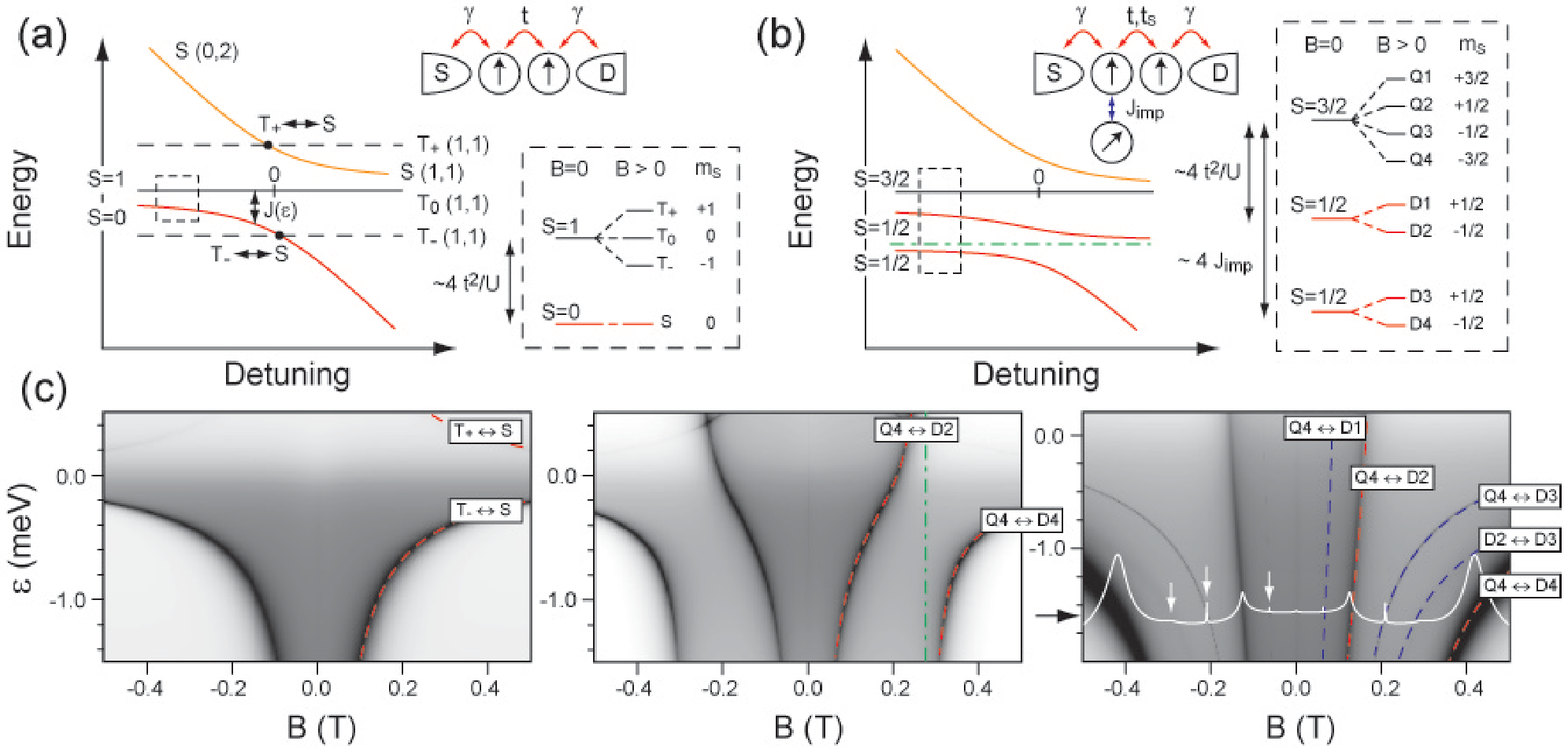}
\caption{\label{Fig3}\textbf{(a)} Energy of the relevant two-electron states of a double quantum dot as a function of detuning. The black line represents
the (1,1) triplet states ($S=1$), while the red and orange curves
represent mixtures of a (0,2) singlet state with a (1,1) singlet state
($S=0$). For non-zero magnetic field the triplet state energies
split. For a given detuning and magnetic field there are points where
the triplet energies equal the singlet energies (marked by dots). \textbf{(b)} The energy
of the relevant two-electron states of a double quantum dot in the
Pauli blockade regime including a spin-1/2 impurity as a function
of detuning. The impurity spin couples via isotropic Heisenberg interaction with a coupling strength
$J_{imp}$ to one of the dot spins, see inset. The black line represents a $S=3/2$ quartet. These
states are blocked and will have high occupation. The remaining curves
are $S=1/2$ doublets. Spin selection rules allow transitions with
$\Delta m_{s}=\pm1$: $Q3 \leftrightarrow D1$, $Q3 \leftrightarrow D3$,
$Q4 \leftrightarrow D2$, $Q4 \leftrightarrow D4$
and $D2 \leftrightarrow D3$, while other, forbidden,
transitions are $Q4 \leftrightarrow D1$ and $Q4 \leftrightarrow D3$. \textbf{(c)} Current as a function of magnetic field
and detuning. Left: for no impurity spin ($J_{imp}=0\,\mu$eV) and $t=70\,\mu$eV, Middle:
for impurity spin with $J_{imp}=8\,\mu$eV, and $t=87\,\mu$eV,
no spin-flip tunneling, Right: for impurity spin with $J_{imp}=6\,\mu$eV, and $t=170\,\mu$eV, in the presence of spin-flip
tunneling with $t_s=7.5\,\mu$eV. Note that the origin of the curve $D2 \leftrightarrow D3$ in the rightmost plot is the large interdot tunnel coupling and not the spin-orbit interaction.}
\end{figure*}

As evident from the measurements presented in Fig.~2(b),
the observed evolution of the spin states can be very rich and, as
discussed below, deviates considerably from naive
expectations based on the simple even-odd spin filling pattern
observed in Fig.~1. We start with a description of the
current dependence at $V_{\textrm{bar}}=0$ mV, corresponding to
the leftmost stability diagram in Fig.~2(a). In the Pauli
blockade regime only the (1,1) triplet states are occupied and no current can pass
through the device. However, when the magnetic field is non-zero the
degenerate triplet splits and for a given magnetic field and
detuning the triplet energies equal the hybridized singlet energies, see schematics in Fig. 3(a).
In the presence of spin relaxation such as due to spin-orbit mediated electron-phonon
interaction \cite{Bulaev}, the spin flip rate is much higher at
these points and provides an efficient escape route from the
blockaded triplet state. Thus measuring the current as a function of detuning and
magnetic field maps out the exact energy dependence of the singlet
energies, and therefore the singlet-triplet exchange energy $J(\epsilon)$, as seen in the leftmost funnel-shaped pattern in
Fig.~2(b). We obtain an excellent fit to the experimental data using a theoretical model of
the device (for details see Methods section) which yields a tunnel coupling $t=70$ $\mu$eV,
consistent with an independent estimate of $t$ from the stability diagram in Fig. 2(a). The measurements also demonstrate our ability to electrically tune $J(\epsilon)$ by varying the detuning which, in the limit of large negative detuning, approaches $J(\epsilon) \sim 4 t^2 /U$, where $U \sim 10-20$ meV is the charging energy of the quantum dots, see schematics in Fig.~3(a). The calculated
current for this situation is shown in the leftmost plot of Fig.~3(c).

While the funnel-shaped pattern in the leftmost panel in Fig.~2(b)
has been observed previously in other systems such as GaAs \cite{Petta}, the evolution of the
spin states in the remaining two panels has
not previously been seen in \textit{any} double quantum dot.
When the barrier voltage is set to $V_{\textrm{bar}}=$10 mV a completely
different pattern appears in which two sets of curves that approach
a common asymptote at $B \sim $ 0.26 T are seen, see dashed-dotted line. Intriguingly, when the barrier
voltage is increased further to $V_{\textrm{bar}}=$15 mV, the two
sets of curves are accompanied by three further, weaker, sets of curves
[see the rightmost panel in Fig.~\ref{Fig2}(b)].

To explain our observations, we propose a model where additionally
to the electrons on the double dot a third spin, the impurity spin,
is present. As justified below, we assume it to be a spin-1/2 that couples to the spin
of one of the dot electrons via an isotropic Heisenberg interaction with a coupling strength $J_{imp}$, see Fig.~3(b). The strength of the
interaction as well as that of the tunnel coupling are tuneable by the gate electrodes. The relevant states of the combined quantum
dot and impurity system can be characterized by their total spin:
A fourfold degenerate spin 3/2 state and two doubly degenerate spin
1/2 states, as shown in Fig.~3(b). The $S=3/2$ states
cannot mix with the $S=1/2$ (0,2) state since tunneling conserves
spin, and therefore block the current, while all the spin 1/2 states can
take part in transport through the device, having a (0,2) component.
The multiplets split in a magnetic field with the energy of states
with higher magnetic spin quantum numbers $m_S$ passing the energy of states
with lower magnetic spin quantum numbers as indicated in Fig.~3(b).
At finite magnetic fields, the lowest lying state in the $S=3/2$ quartet ($Q4$)
has the highest occupation probability. Since selection rules
limit the possible transitions to those from state $Q4$ to $D2$ and $D4$, we therefore expect to see two strong
curves, tracing out the shape of the $S=1/2$ levels. This is indeed the case as observed in
the data, see Fig. 2(b), which is in excellent agreement with the calculated current as shown in Fig.3(c).

The model also allows us to investigate the effects of spin-orbit interaction on the electron transitions. A first indication of the presence of spin-orbit coupling is the zero-field dip in the spin blockade leakage current around $\epsilon = 0$ as seen in the leftmost plot in Fig. 2(b). This feature has previously been observed in carbon nanotube \cite{Churchill1} and InAs \cite{Pfund1} double quantum dots and has been tentatively attributed to spin-orbit interaction. These results were reproduced in recent theoretical work in which spin-orbit interaction was shown to introduce non spin-conserving tunneling between the two quantum dots \cite{Nazarov}. In our model this is characterized by a spin-flip tunneling amplitude $t_s$. Since in the presence of spin-orbit interaction $m_s$ is no longer a good quantum number, spin selection rules can be violated, resulting in additional transitions and therefore extra curves in the current plots. These curves are indeed observed in the data, as most clearly seen in the rightmost plot in Fig. 2(b). The position and evolution of the additional three curves are in excellent agreement with those predicted by the model calculation as illustrated by the dotted curves in Fig. 2(b) and the corresponding calculated current in Fig. 3(c).

Spin-orbit interaction also affects the shape of the curves. The spin-flip tunneling induces coherent transitions between T$_{+/-}$(1,1) and S(0,2) resulting in an avoided crossing between the triplet and singlet states, the size of which depends on the (0,2) component of the singlet. This results in relatively narrow inner and a wide outermost curve which increases in width as $\epsilon \rightarrow 0$. A comparison with the data gives fairly narrow bounds for the spin-flip tunneling and yields $t_s = 7.5$ $\mu$eV, see Supplementary Information. Using the estimate from Ref.~\cite{Nazarov} and an orbital energy $E_{orb}$ of the order of 2 meV we can deduce the spin-orbit interaction energy as $E_{SO}=E_{orb} t_{s}/t \approx$ 0.1 meV. We note that this value is similar to previous estimates \cite{Kuemmeth, Churchill2} even though the strength of the spin-orbit interaction is not \textit{a priori} clear in the many-electron limit and for mixed orbital states.

The above analysis demonstrates that we are able to detect the presence of a single impurity spin coupled to the carbon nanotube double quantum dot and determine its spin quantum number. An important question that remains is the nature of the impurity. Even a single magnetic atom absorbed on the nanotube wall would have a significant effect on the spin states \cite{Odom} and could be present as a result of the nanotube growth process, see Methods section. Further possibilities are charge traps in the gate oxide or the presence of a defect such as a vacancy or dopant in the carbon lattice. Even high-quality carbon nanotubes have been found to contain one defect per 4 $\mu$m on average \cite{Fan}. This would imply that the majority of our devices (having total length $L \sim 1 \mu$m) contain either zero or one impurity. Whereas many nanotube devices will therefore be without a single defect, requirements on the nanotube quality will become more stringent when many quantum dots are coupled in a large-scale quantum circuit.

Our experiments also show that the exchange interaction between the carbon nanotube and impurity spin can be precisely controlled with a gate electrode, see Fig. 2(b). In combination with the recent advances in attaching single atoms or molecules to the nanotube sidewall \cite{Banerjee}, this suggests the possibility of storing quantum information into the attached groups and using the carbon nanotube as a quantum bus and for spin state read-out \cite{Elzerman, Wabnig}. The ability to control these interactions will be instrumental in developing carbon materials for quantum information processing.


\appendix

\section{Methods}

\subsection{Device fabrication}

Carbon nanotubes were synthesized by chemical vapor deposition using a procedure similar to Ref. \cite{Kong}. To form the catalyst particles, $\sim 0.5$  $\mu$g/cm$^{3}$ Fe(NO$_3$).9H$_2$O (Sigma-Aldrich) was sonicated in isopropanol. The bare Si/SiO$_2$ substrate (300 nm thermal oxide) was dipped into the solution immediately after sonication and dried by air blowing. It was then heated to 900 $^o$C in a 1" tube furnace under a hydrogen gas flow of 400 sccm during which the iron nitrate was reduced to iron particles. Once at this temperature, 500 sccm of methane (containing natural isotope ratios) was added to the flow for 8 minutes. After the methane flow was stopped, the substrate was cooled under the same 400 sccm hydrogen flow. Alignment marks were defined by electron-beam lithography and scanning electron microscopy and atomic force microscopy were used to locate the nanotubes in relation to these. Further steps of electron-beam lithography and evaporation contact the nanotubes with 20 nm thick, 300 nm wide gold ohmic contacts, separated by 700 nm, which also form the outer barriers of the quantum dots. The final electron-beam lithography layer writes the 100 nm wide and 100 nm spaced gates, which are made from two 1.2 nm layers of aluminum, oxidized in air and capped with a further 20 nm of titanium and 6 nm of gold. The source lead is grounded by the virtual earth of a $\times 10^7$ current to voltage preamplifier and the drain has a small variable dc bias applied. The chip is mounted on the cold finger of a 60 mK dilution refrigerator and connected to the measurement apparatus with filtered low-frequency lines.

\subsection{Theoretical model}

We use a master equation technique to model the physical system of
interest that consists of a double quantum dot coupled to an impurity
spin. We are interested in the electrical current through the double quantum dot
in the presence of spin relaxation and decoherence.

The dynamics of the electrons on the double quantum dot is modeled
by an extended two-site Hubbard Hamiltonian with single orbitals.
The impurity spin is modeled as a spin-1/2 coupled to one of the
dots via an isotropic Heisenberg interaction. To facilitate the
numerical evaluations we represent the density matrix of the
system in a many-body basis resulting in 32 basis states. The
calculations are performed in a rotated basis in which the
Hamiltonian of the system is diagonal.

The double quantum dot is connected to left and right metallic leads which are described as
reservoirs of non-interacting fermions. The applied magnetic field
does not affect the distribution function of the fermions, as long
as the Zeeman energy is small compared to the Fermi energy. The
tunneling between the leads and the double quantum dot is treated
perturbatively to second order in the tunneling amplitude. To
model spin relaxation and decoherence each spin is coupled to an
independent bath of harmonic oscillators in thermal equilibrium
with a flat spectral density. The coupling between the spin and
the bath is treated perturbatively to lowest non-vanishing order
in the coupling strength.

Within the Born and Markov approximations we derive an equation of motion for the density
matrix of the double quantum dot and the impurity spin, that we
solve for the stationary state. Using a similar methodology we
derive the current operator up to second order in the tunneling
amplitude and calculate the electrical current in the stationary
state. Further details of the model can be found in the Supplementary Information.

\subsection{Acknowledgements}

We thank David Cobden and Jiang Wei for the carbon nanotube growth, and Brendon Lovett for discussions. This research was supported by EPSRC through
QIP IRC (GR/S82176/01). J.W. thanks the Wenner-Gren Foundations for
financial support. M.R.B. acknowledges support from the Royal Society.

\onecolumngrid
\newpage

\setcounter{page}{1}
\thispagestyle{empty}

\begin{doublespace}

\begin{center}
\textbf{{\large Supplementary Information for\\
"Single Spin Detection with a Carbon Nanotube Double Quantum Dot"}}\\
\bigskip
S. J. Chorley, G. Giavaras, J. Wabnig, G.A.C. Jones, C.G. Smith, G.A.D. Briggs, and M.R. Buitelaar\\

\textit{Cavendish Laboratory, University of Cambridge, JJ Thomson Avenue, Cambridge, CB3 0HE, United Kingdom}

\textit{Department of Materials, University of Oxford, Parks Road, Oxford,
OX1 3PH, United Kingdom}

\end{center}

\setcounter{figure}{0}

\renewcommand{\figurename}{Figure S} 


\section{Experimental data}

In the following sections we present additional data to that in the main text. Figures S1(a,b) show stability diagrams of the device used to obtain the various capacitance terms and charging energies of the two quantum dots following the work of van der Wiel \textit{et al} \cite{Wiel}. The parameters are summarized in Table I. The difference in the charging energies $U$, as well as the single-particle energies $\Delta E$  might be related to geometry: QD$_R$ is slightly smaller than QD$_L$, see inset in Fig. S1(a) which shows an SEM image of the device taken after measurement.

\begin{table}[b]
\begin{center}

\begin{doublespace}
\caption{Capacitance terms and charging energies of the double quantum dot}
\end{doublespace}

\begin{tabular}{p{2cm} p{1.2cm} p{1.2cm} p{1.2cm} l} \\\hline\hline
\textbf{Property} & \textbf{QD$_{\textrm{L}}$} & \textbf{QD$_{\textrm{R}}$} & \textbf{Unit} & \textbf{Comment} \\\hline
$C$ & 15 & 6.2 & aF & Quantum dot capacitance \\
$C_{m}$ & \multicolumn{2}{l}{~~~~0.22} & aF & Interdot capacitance \\
$C_{g}$ & 0.12 & 0.060 & aF & Plunger gate-QD capacitance \\
$U$ & 11 & 26 & meV & Charging energy \\
$U'$ & \multicolumn{2}{l}{~~~~0.37} & meV & Electrostatic coupling energy \\
$\Delta E$ & 1.8 & 2.0 & meV & Single-particle addition energy
\\\hline
\end{tabular}
\end{center}
\end{table}

\begin{figure*}
\includegraphics[width=140mm]{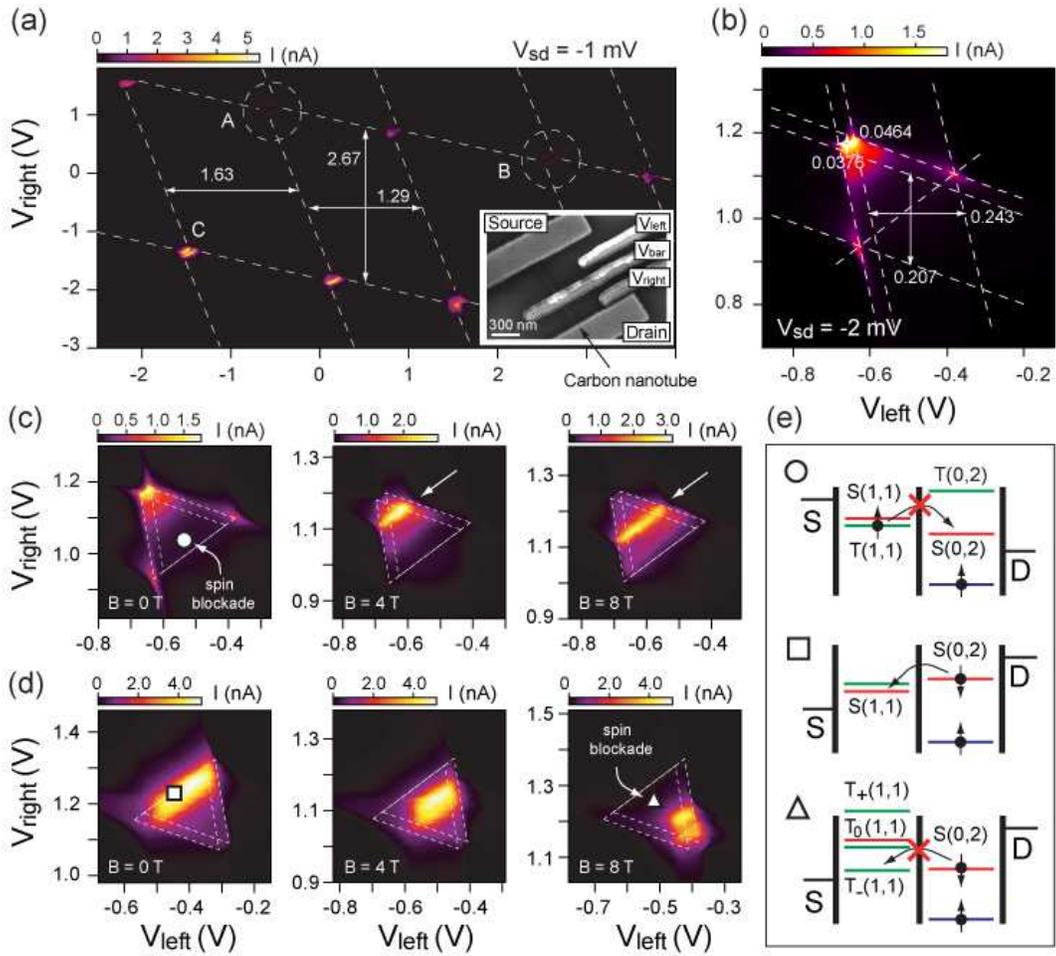}
\begin{doublespace}
\caption{\label{SI_Fig1}\textbf{(a)} Charge stability diagram for $V_{sd} = -1$ mV. The arrows indicate the spacing in $V_{left/right}$ of the hexagonal pattern. Inset: Scanning electron micrograph of the device. The arrow indicates the position of the carbon nanotube. \textbf{(b)} Detailed measurement of the bias triangles at the (0,1) to (1,2) charge transition (effective electron occupancies) for $V_{\textrm{sd}}=-2$ mV. \textbf{(c)} Bias triangles for $V_{\textrm{sd}}=-2$ mV and $B$ = 0, 4, and 8 T. The arrow in the measurements at $B=4,8$ T indicates transitions to the $T(0,2)$ state. \textbf{(d)} Bias triangles for $V_{\textrm{sd}}=+2$ mV and $B$ = 0, 4, and 8 T. Spin blockade is observed at large magnetic fields. \textbf{(e)} Energy schematics of the double quantum dot. The symbols correspond to the positions in the bias triangles in panels (c,d).}
\end{doublespace}
\end{figure*}

Figures S1(c,d) show measurements for various magnetic fields ($B= 0,4,8$ T) perpendicular to the nanotube axis at both negative and positive applied bias. These measurements provide additional evidence for the interpretation of the data in terms of spin blockade and allow us to identify excited states in the bias window. Fig. S1(c) shows that at finite field the current at the lower edges of the spin blockade triangles are suppressed as expected when the triplet states split off and spin exchange with the leads is suppressed. The spin-blockade region also becomes smaller as the $T_-$ states move down in energy with field and transitions to $T(0,2)$, indicated by the arrows, are observed at smaller detuning. As shown in Fig. S1(d), the behavior is precisely the opposite for positive ($V_{sd} = + 2$ mV) bias voltages. At $B=0$ T the double dot is not spin blockaded and a large current at the base of the triangles is observed. However, as the field increases, the double dot enters the spin blockade regime as the $T_-$ triplet pushes the base of the triangle out, but blocks the $S(0,2) \rightarrow T_-(1,1)$ transition, see schematic in Fig. S1(e).

\begin{figure*}
\includegraphics[width=100mm]{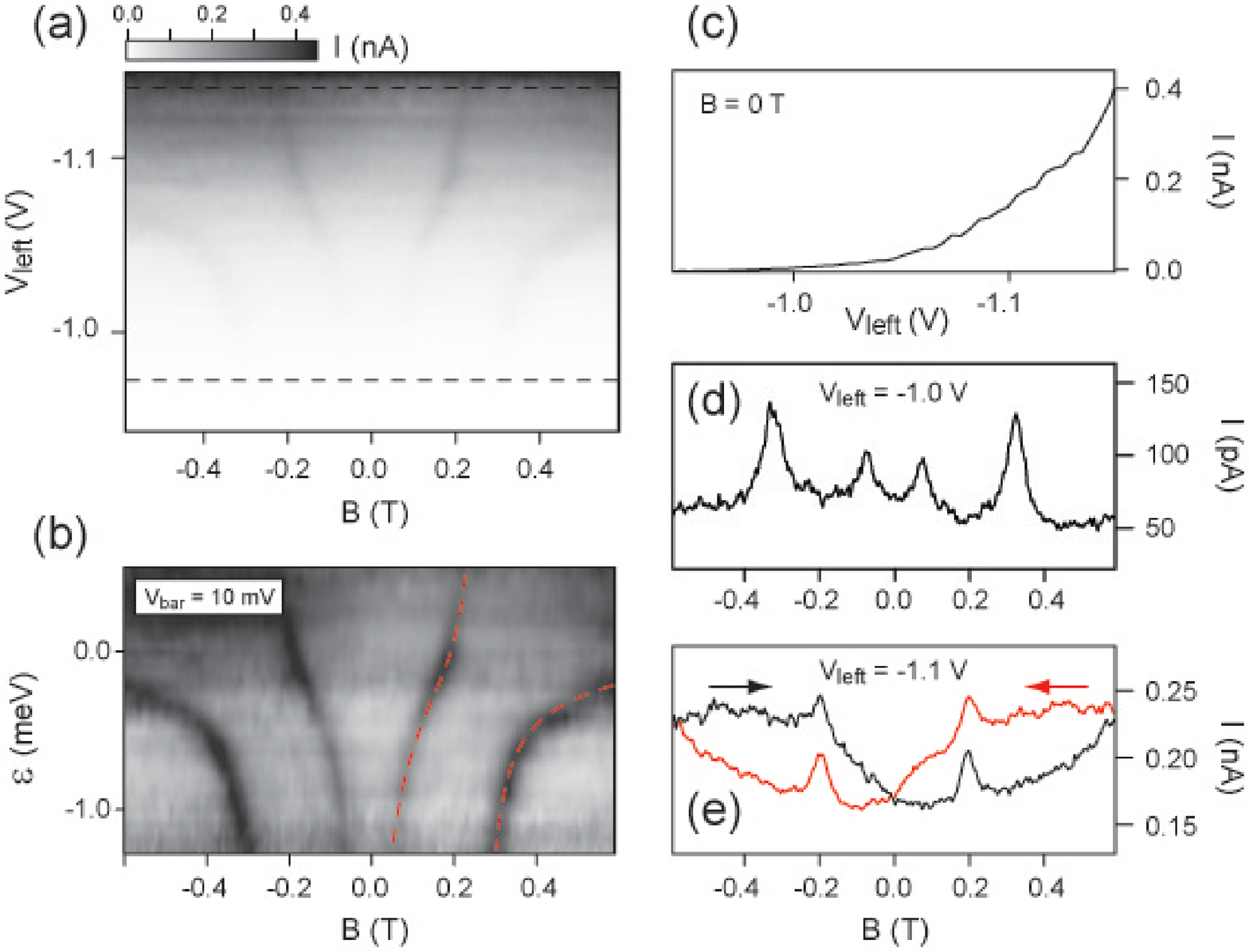}
\begin{doublespace}
\caption{\label{SI_Fig2}\textbf{(a)} Current as-measured as a function of detuning (projected on $V_{left}$) and magnetic field. \textbf{(b)} Normalized current corresponding to the section of panel (a) between the dashed lines. The dashed red curves are fits to the theory, see main text. \textbf{(c)} Current as function of detuning (projected on $V_{left}$) at $B=0$ T. \textbf{(d)} Current as function of magnetic field for $V_{left}=-1.0$ V. \textbf{(e)} Current as function of magnetic field for $V_{left} = -1.1$ V for both sweep directions, as indicated by the arrows.}
\end{doublespace}
\end{figure*}

Figure S2 illustrates how the normalized data as presented in the main text is obtained from the current as-measured. Fig. S2(a) shows the current as a function of detuning and magnetic field for spin-blockaded bias triangles at $V_{sd} = -1$ mV and $V_{bar} = 10$ mV, see Fig. 2 in main text. As shown in Fig. S2(c), the current varies strongly as a function of detuning. The corresponding normalized current, shown in Fig. S2(b), accentuates the evolution of the spin states by effectively subtracting the background current which allows for a detailed comparison with theory (dashed red lines). The values on the detuning axis can be obtained from the stability diagrams, as well as from fits to the theory. Both methods give very similar estimates for the detuning. In the measurements presented here, the values on the detuning axes are those obtained from fits to the theory.

We have observed a dependence of the current measurements on the direction and rate of the magnetic field sweeps. This is most clearly seen for positive detuning, as illustrated in Fig. S2(e) for a sweep rate of 20 T/h. Note, however, that the position of the current peaks in magnetic field appears to be unaffected. As a nearby Ge-thermometer showed similar behavior we attribute the observed hysteresis to an artefact of the measurement system (possibly heating) and not to coupling of the electrons to the nuclear spin system (our nanotubes contain $\sim 1 \% ~^{13}$C).

\begin{figure*}
\includegraphics[width=155mm]{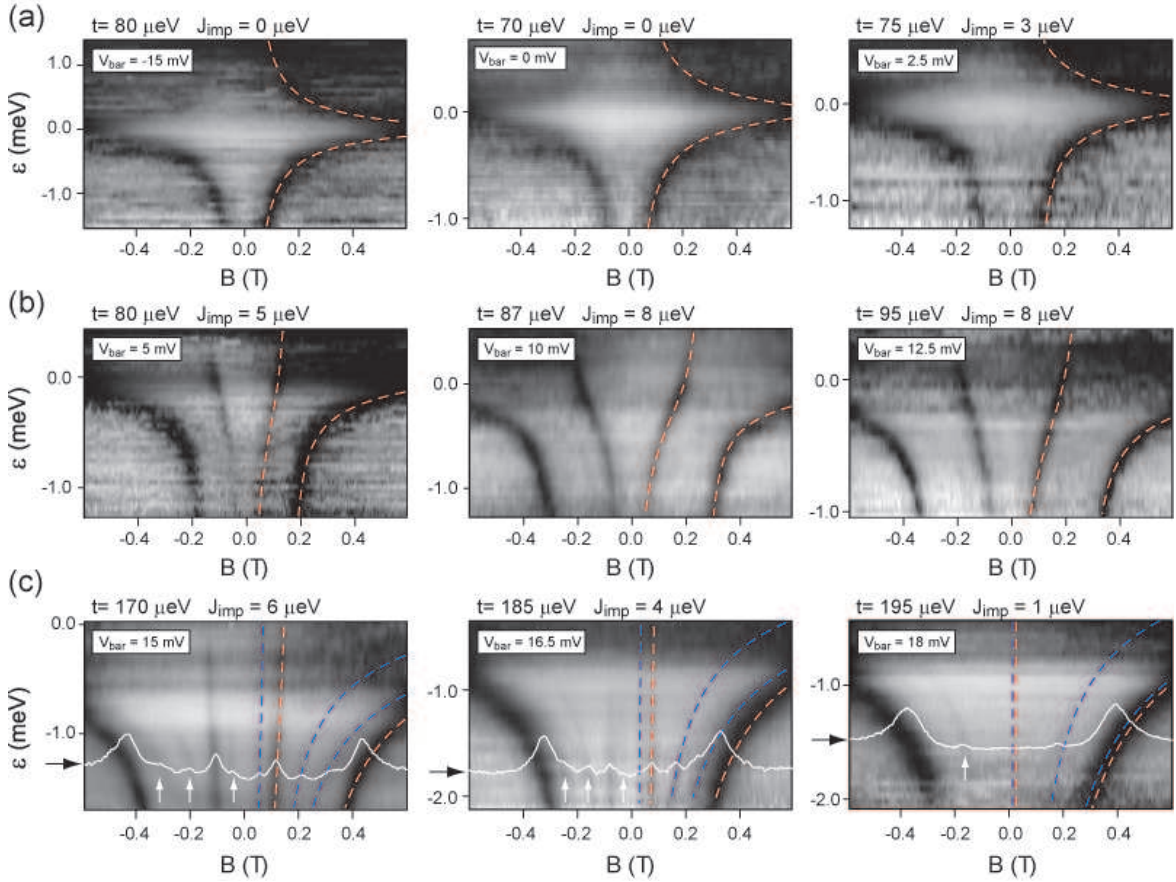}
\begin{doublespace}
\caption{\label{SI_Fig3}\textbf{(a-c)} Measurements of the normalized current as a function of magnetic field and detuning for nine different $V_{bar}$. In panel (a) the exchange interaction $J_{imp}$ is small and only a single set of lines is visible. In panel (b) the exchange interaction is stronger and an additional set of lines appears. The evolution of the lines can be fitted very well with a theoretical model of the device, see dashed red lines. For the measurements shown in panel (c), the tunnel coupling between the two quantum dots is relatively strong and three additional, faintly visible, sets of lines appear, as indicated by the arrows. As described in the main text, the evolution of these lines follows that predicted by the theoretical model, see dashed blue lines.}
\end{doublespace}
\end{figure*}

\begin{figure*}
\includegraphics[width=120mm]{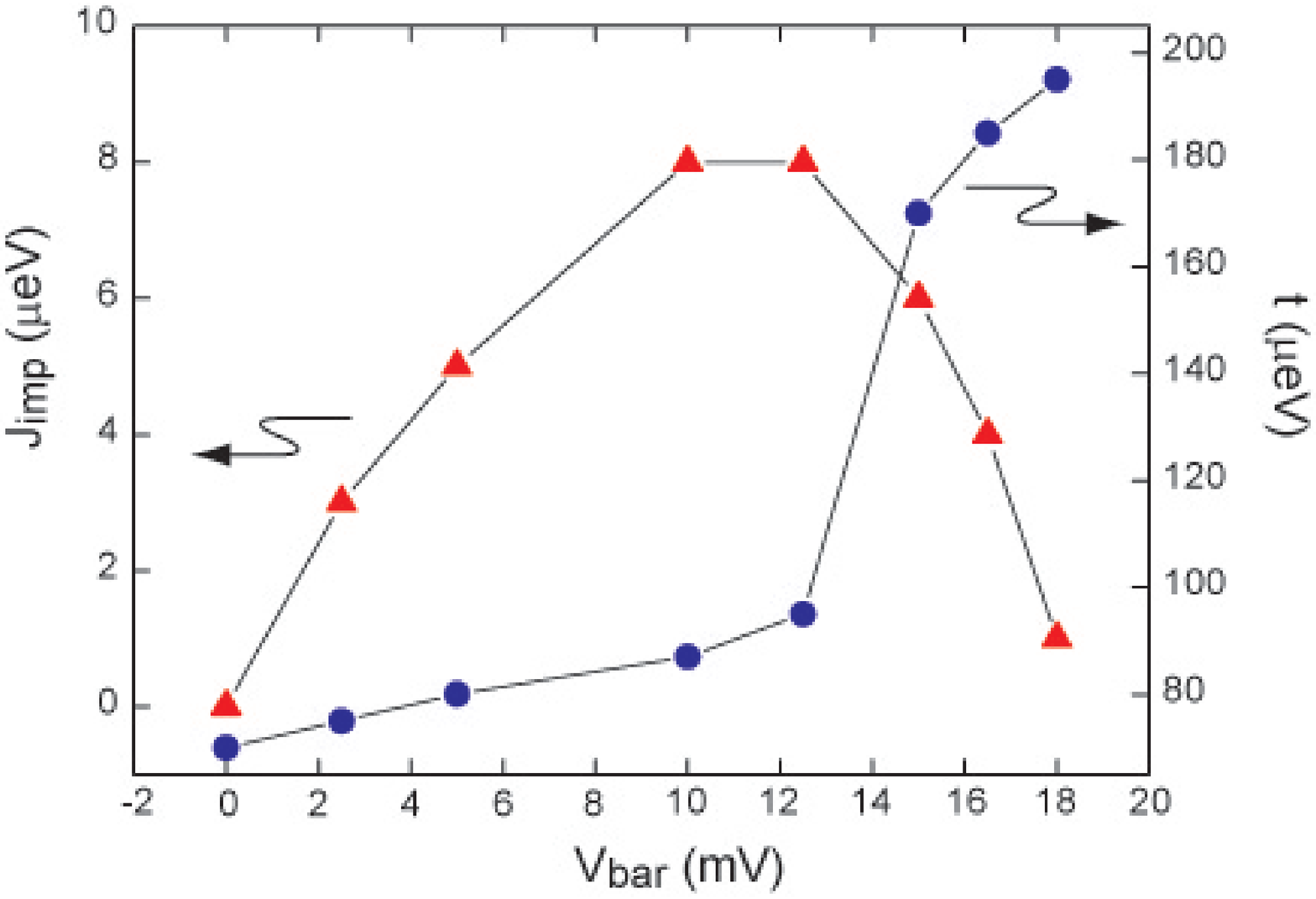}
\begin{doublespace}
\caption{\label{SI_Fig4} Evolution of the quantum dot-impurity coupling strength $J_{imp}$ (left) and tunnel coupling $t$ (right) with barrier gate voltage $V_{bar}$.}
\end{doublespace}
\end{figure*}

Figure S3 shows the complete normalized data set, measured at $V_{sd} = -1$ mV for nine different $V_{bar}$, for the spin-blockaded bias triangles labeled $A$ in Fig. S1(a). The coupling to the impurity spin is small in Fig.~S3(a) but is clearly observed in Figs.~S3(b,c). For relatively large $t$, additional curves (dashed blue lines) become faintly visible, see Fig.~S3(c). All measurements allow for a detailed fit to the theory with the results for $t$ and $J_{imp}$ summarized in Fig.~S4. Both the tunnel coupling between the quantum dots and the coupling to the impurity spin strongly depend on - that is, are tuneable by - the barrier gate electrode. The tunnel coupling increases for increasing $V_{bar}$, most strongly so after $V_{bar} = 12.5$ mV while the coupling to the impurity spin increases and then decreases again.

\begin{figure*}
\includegraphics[width=100mm]{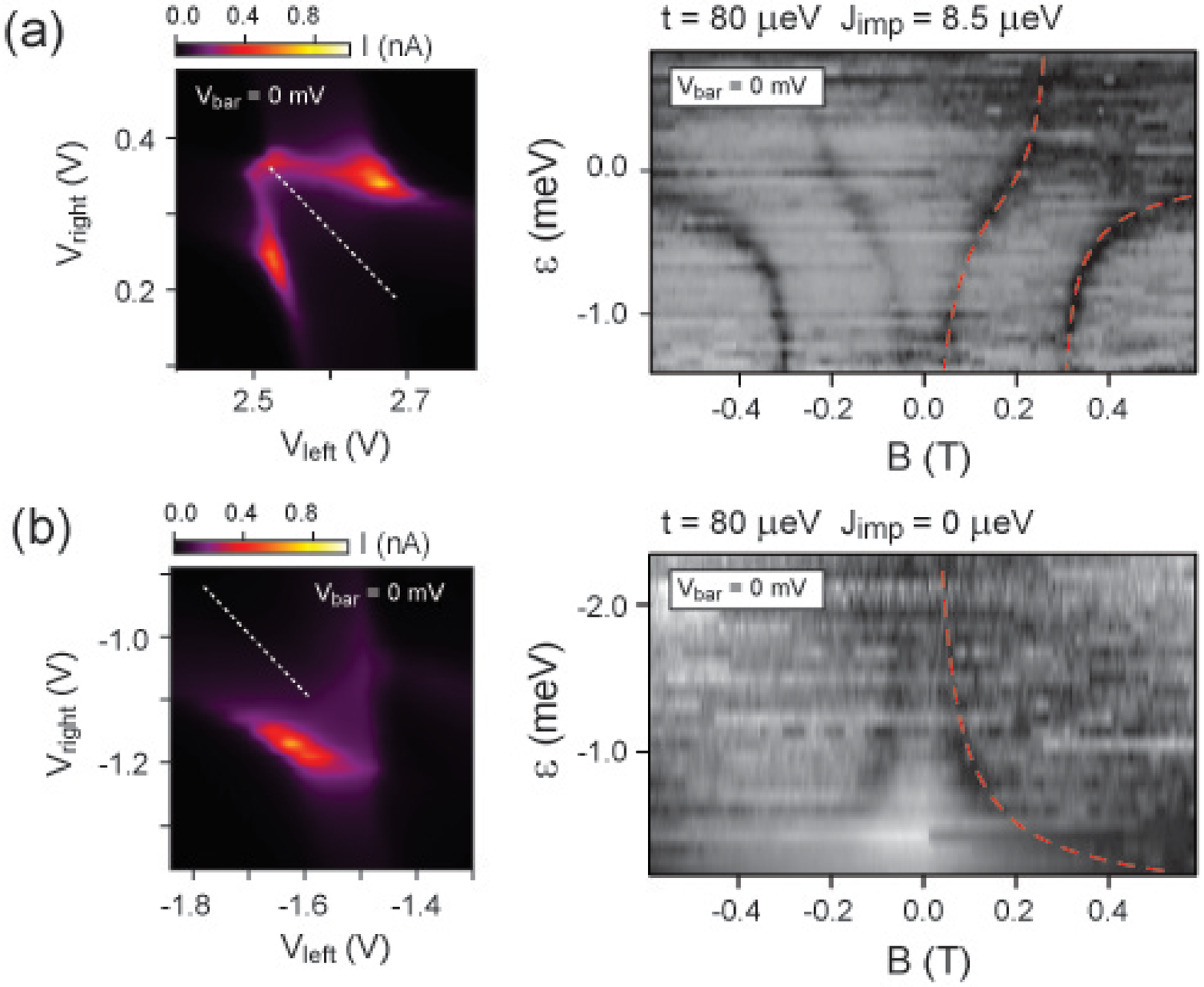}
\begin{doublespace}
\caption{\label{SI_Fig5}\textbf{(a,b)} Spin-blockaded bias triangles and spin-state evolution as a function of detuning and magnetic field for two spin configurations different from those discussed in the main text. The dashed red lines are fits to the theory.}
\end{doublespace}
\end{figure*}

Finally, in Fig. S5 we show measurements on two sets of spin-blockaded bias triangles different from those discussed in the main text. The stability diagrams in Fig. S5(a,b), both measured for $V_{bar} = 0$ mV, correspond to the regions labeled $B$ and $C$ in Fig. S1(a), respectively. The results are very similar to those discussed in the main text, demonstrating that our results are reproducible and that, apart from the parity of the electron occupation (even or odd), the qualitative features of the measurements do not depend on the total number of electrons on the quantum dots. We furthermore verified that the observed features and the evolution of the spin states with detuning and magnetic field do not depend on the applied bias (tested for $V_{sd} = -0.5, -1, -2, -3, -5$ mV). As expected, the features are not observed for non-spin blockaded bias triangles. Also note that in Fig. S5(a), we observe a strong coupling of the electrons on the quantum dots to the impurity spin ($J_{imp} = 8.5$ $\mu$eV), even for $V_{bar} = 0$ mV. This indicates that the coupling to the impurity spin is tuneable not only by the barrier gate electrode (as shown in the main text), but also by the side gates.

\clearpage

\section{Theoretical Model and results}

In this section we describe in detail the theoretical model and
present results for the electrical current for some special cases.
The total system consists of a double quantum dot, a nearby
impurity spin-1/2, the metallic leads, and bosonic heat baths.
This system is modelled by the Hamiltonian
\begin{equation}
H_{tot}=H_{S}+H_{leads}+H_{T}+H_{B}+H_{SB}.
\end{equation}
The system Hamiltonian, $H_{S}$, models the double quantum dot and
the impurity spin, $H_{leads}$ ($H_{B}$) models the leads (heat
baths) and $H_{T}$ ($H_{SB}$) models the interaction between the
leads (heat baths) and the system. Specifically, we have
\begin{equation}
H_{S}=H_{DD}+H_{M}+H_{I},\label{system}
\end{equation}
where $H_{DD}$ is a Hubbard Hamiltonian describing the double
quantum dot, $H_{M}$ is due to the applied magnetic field and
$H_{I}$ models the interaction of the double quantum dot with the
impurity spin. For the double quantum dot we have
\begin{equation}
H_{DD}=\sum_{i=1}^{2}\varepsilon_{i}n_{i}-\sum_{\sigma
,\sigma'}t_{\sigma\sigma'}(c_{1\sigma}^{\dagger}c_{2\sigma'}+c_{2\sigma'}^{\dagger}c_{1\sigma})
+\sum_{i=1}^{2}U_{i}n_{i\uparrow}n_{i\downarrow}+U'n_{1}n_{2},
\end{equation}
that allows up to two electrons per dot. The number operator is
$n_{i}=\sum_{\sigma}n_{i\sigma}=\sum_{\sigma}c_{i\sigma}^{\dagger}c_{i\sigma}$
for dot $i=\{1,2\}$ and spin $\sigma=\{\uparrow,\downarrow\}$. The
operator $c_{i\sigma}^{\dagger}$ ($c_{i\sigma}$) creates
(annihilates) an electron on dot $i$ with on-site energy
$\varepsilon_{i}$. The spin conserving tunnel coupling between the
two dots is $t_{\sigma\sigma'}=t$ for $\sigma=\sigma'$, and the
spin-flip tunnel coupling between the two dots is
$t_{\sigma\sigma'}=t_{s}$ for $\sigma\neq\sigma'$. $U_{i}$ is the
charging energy of each dot and $U'$ is the interdot Coulomb energy
between the dots. The Hamiltonian part due to the applied magnetic
field is
\begin{equation}
H_{M}=\sum_{i=0}^{2}\frac{\Delta_{i}}{2}\sigma_{i}^{z},
\end{equation}
where $i=0$ refers to the impurity spin and the spin operators are
defined in the standard way
$\bm{\sigma}_{i}=\sum_{\sigma\sigma'}c_{i\sigma}^{\dagger}\bm{\sigma}_{\sigma\sigma'}c_{i\sigma'}$,
where $\bm{\sigma}$ is the vector of the $2\times2$ Pauli
matrices. $\Delta_{i}=g_{i}\mu_{B}B$ is the Zeeman splitting due
to the magnetic field $B$ along $z$, $g$-factor $g_{i}$ and the
Bohr magneton $\mu_{B}$. We consider a Heisenberg interaction
between dot 1 ($i=1$) and the nearby impurity spin ($i=0$) of the
general form
\begin{equation}
H_{I}=J_{x}\sigma_{0}^{x}\sigma_{1}^{x}+J_{y}\sigma_{0}^{y}\sigma_{1}^{y}+J_{z}\sigma_{0}^{z}\sigma_{1}^{z},\label{inter}
\end{equation}
with $J_{x}$, $J_{y}$, $J_{z}$being the strength of the
interaction, and we set $J_{x}=J_{y}=J_{z}=J_{imp}$ for isotropic
interaction.

Having described the system of interest, i.e., the double quantum
dot and the impurity spin-1/2, we turn to the metallic leads, the
bosonic heat baths and their interaction with the system. The left
and right leads are described by a Hamiltonian of the form
\begin{equation}
H_{leads}=\sum_{\ell k\sigma}\epsilon_{\ell k}d_{\ell
k\sigma}^{\dagger}d_{\ell k\sigma},
\end{equation}
where $d_{\ell k\sigma}^{\dagger}$ ($d_{\ell k\sigma}$) creates
(annihilates) an electron in lead $\ell=\{L,R\}$ with momentum
$k$, spin $\sigma$ and energy $\epsilon_{\ell k}$. The interaction
between the dots and the leads is given by the tunneling
Hamiltonian
\begin{equation}
H_{T}=\sum_{k\sigma}(t_{L}c_{1\sigma}^{\dagger}d_{Lk\sigma}+t_{R}c_{2\sigma}^{\dagger}d_{Rk\sigma})+\text{H.c.},
\end{equation}
where $t_{L}(t_{R})$ is the tunnel coupling between dot $1(2)$ and
lead $L(R)$ and we consider the symmetric case where
$t_{L}=t_{R}$. To include spin relaxation and decoherence we
consider a generic bosonic heat bath that is modelled as a set of
harmonic oscillators and is described by the Hamiltonian
\begin{equation}
H_{B}=\sum^{2}_{i=0}\sum_{j}\hbar\omega_{i,j}a_{i,j}^{\dagger}a_{i,j}.
\end{equation}
We assume that the impurity spin and each dot are coupled to an
independent bosonic heat bath ($i=0$ refers to the impurity spin
bath and $i=1$, 2 to dot 1, 2) and there are no
environment-induced correlations between them. Moreover, we assume
that the spins are flipped one at a time. The operator
$a_{i,j}^{\dagger}$ ($a_{i,j}$) creates (annihilates) a boson in
mode $j$ and $\omega_{i,j}$ are the frequencies of the bath modes.
The impurity spin and the spins on the double quantum dot interact
with the corresponding bath via the general model Hamiltonian
\begin{equation}
H_{SB}=\sum_{i=0}^{2}\sigma_{i}^{-}\sum_{j}\Lambda_{i,j}a_{i,j}^{\dagger}+\text{H.c.},\label{bathint}
\end{equation}
where the spin-flip operators are
$\sigma_{i}^{-}=c_{i\downarrow}^{\dagger}c_{i\uparrow}$ and
$\Lambda_{i,j}$ is the coupling constant between the impurity spin
$(i=0)$, dot 1 $(i=1)$ or dot 2 $(i=2)$ and the $j$th mode of the
corresponding bath. $H_{SB}$ allows spin-flip processes for the
impurity spin and the spins on the double quantum dot, via energy
exchange with the heat bath, which lead to a leakage current in
the spin blockade regime.

To investigate the electronic transport through the double quantum
dot we employ a master equation approach and derive an equation of
motion for the reduced density matrix, $\rho$, for the system of
interest (double quantum dot and impurity spin). Starting with the
density matrix of the total system, $\chi_{tot}$, and within the
standard Born and Markov approximations we derive an equation of
motion for $\rho$ by tracing over the leads and bosonic baths
degrees of freedom, i.e., $\rho$=Tr$_{E}\{\chi_{tot}\}$ where
Tr$_{E}\{...\}$ means trace over the environmental degrees of
freedom. It can be shown that the equation of motion of the
density matrix $\rho$ can be written in a compact form with the
superoperators $\mathcal{L}_{S}$, $\mathcal{L}_{leads}$ and
$\mathcal{L}_{B}$ as
\begin{equation}
\dot{\rho}(t)=\mathcal{L}_{S}\rho(t)+\mathcal{L}_{leads}\rho(t)+\mathcal{L}_{B}\rho(t),\label{master}
\end{equation}
with the free evolution term
\begin{equation}
\mathcal{L}_{S}\rho(t)=-\frac{i}{\hbar}[H_{S},\rho(t)]\notag,
\end{equation}
and the terms due to the electronic leads
\begin{equation}
\mathcal{L}_{leads}\rho(t)=
-\frac{1}{\hbar^{2}}\text{Tr}_{E}\{\int_{0}^{\infty}d\tau[H_{T},[U(\tau)H_{T}U^{\dagger}(\tau),\rho(t)\otimes\rho_{leads}]]\}\notag,
\end{equation}
and the bosonic baths
\begin{equation}
\mathcal{L}_{B}\rho(t)=
-\frac{1}{\hbar^{2}}\text{Tr}_{E}\{\int_{0}^{\infty}d\tau[H_{SB},[V(\tau)H_{SB}V^{\dagger}(\tau),\rho(t)\otimes\rho_{B}]]\}.\notag
\end{equation}
Both $H_{T}$ and $H_{SB}$ are treated in second order perturbation
theory with the operators
$U(\tau)=\exp[-i(H_{S}+H_{leads})\tau/\hbar]$ and
$V(\tau)=\exp[-i(H_{S}+H_{B})\tau/\hbar]$, and $\rho_{leads}$,
$\rho_{B}$ being the equilibrium density matrix for the leads and
the bosonic baths respectively which have the same temperature.

The quantity of interest is the electrical current that flows
through the double quantum dot. By definition the current
operator, for example, for the right lead is
\begin{equation}
I_{R}=
-e\dot{N}_{R}=-e\frac{i}{\hbar}[H_{T},N_{R}]=e\frac{i}{\hbar}\sum_{k\sigma}(t_{R}c_{2\sigma}^{\dagger}d_{Rk\sigma}-t^{*}_{R}d_{Rk\sigma}^{\dagger}c_{2\sigma}),
\end{equation}
where $N_{R}=\sum_{k\sigma}d_{Rk\sigma}^{\dagger}d_{Rk\sigma}$ is
the number operator for electrons in the right lead. Tracing out
the leads we derive the expectation value of the current, $\langle
I_{R}\rangle$, that is calculated in the stationary state, i.e.,
when $\dot{\rho}=0$ in Eq.(\ref{master}), in the sequential
tunneling regime. The left lead current is $\langle
I_{L}\rangle$=-$\langle I_{R}\rangle$. For the double quantum dot
with single orbital levels there are sixteen possible states and
the maximum number of electrons is four. Including the two
possible impurity spin states results in a total of thirty two
basis states. For the calculations we write Eq.(\ref{master}) as a
set of coupled equations for all the matrix elements of $\rho$, in
the energy basis, including the normalisation condition for the
diagonal elements that express the occupation probabilities. The
resulting system, for $\dot{\rho}=0$, is then solved numerically. The theoretical model predicts
for the current a rather small increase in the range $|\Delta_i|\lesssim k_B T /4$, due to spin relaxation,
and this can be understood through the Boltzmann factor that arises naturally within the model.

\begin{figure}

\includegraphics[width=165mm]{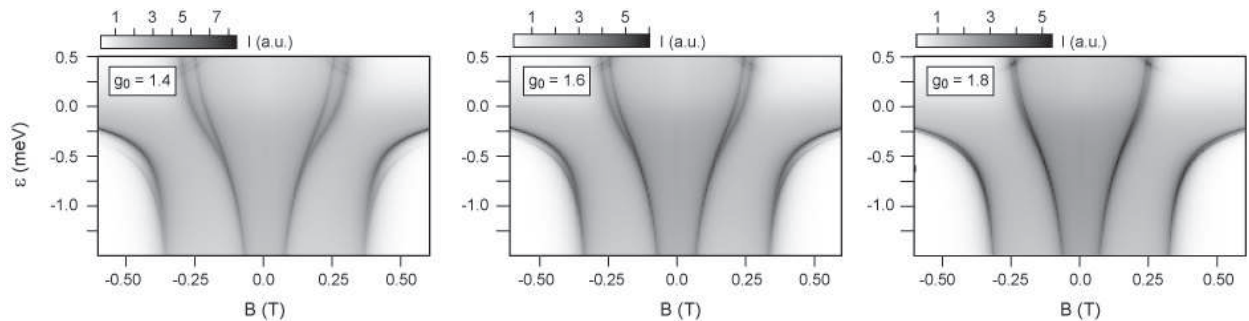}
\begin{doublespace}
\caption{\label{figS1}As Fig.3(c)(middle) in the main article but for different
$g$-factors for the impurity spin and $g_{1}=g_{2}=2$ for the two dots.}
\end{doublespace}
\end{figure}

Finally, we examine the electrical current as a function of
magnetic field and energy detuning for some interesting cases.
Figure~S\ref{figS1} shows the current when the impurity spin and
the double dot spins have different $g$-factors. As the difference
increases the curves of high current begin to split and this
effect becomes noticeable at about $g_{0}=0.8g_{1}$, with
$g_{1}=g_{2}=2$ for the two dots. The exact pattern of the curves
can be identified through the dependence of the three-spin energy
spectrum on magnetic field and detuning when the double dot is in
the spin blockade regime (see also Fig. 3(b) in the main article).
In order to reproduce the experimental data with the theoretical
model the difference in the $g$-factors should be less than about
20\%.

\begin{figure}

\includegraphics[width=165mm]{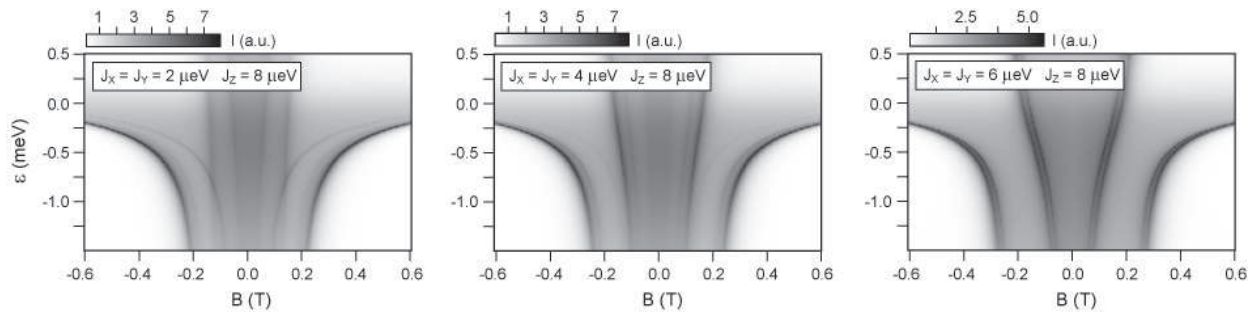}
\begin{doublespace}
\caption{\label{figS2}As Fig.~3(c)(middle) in the main article but for
anisotropic Heisenberg interaction between the impurity spin and one of the quantum dots.}
\end{doublespace}
\end{figure}

Figure~S\ref{figS2} shows the current for anisotropic Heisenberg
interaction between the impurity spin and one of the dots. We
consider antiferromagnetic coupling ($J_{x}$, $J_{y}$, $J_{z}>0$),
which produces a pattern for the high current curves consistent
with the experimental data. Ferromagnetic coupling ($J_{x}$,
$J_{y}$, $J_{z}<0$) produces, in general, a different pattern. As
the anisotropy increases additional curves appear, similar to the
case of spin-flip interdot tunneling described in the main article
(see Fig. 3(c)). However, the analysis of the functional
dependence of the curves on magnetic field and detuning shows
that it is impossible to fit the theoretical results of high
anisotropy to the experimental data. Our calculations suggest that
the anisotropy has to be much smaller than 20\%.

\begin{figure}

\includegraphics[width=165mm]{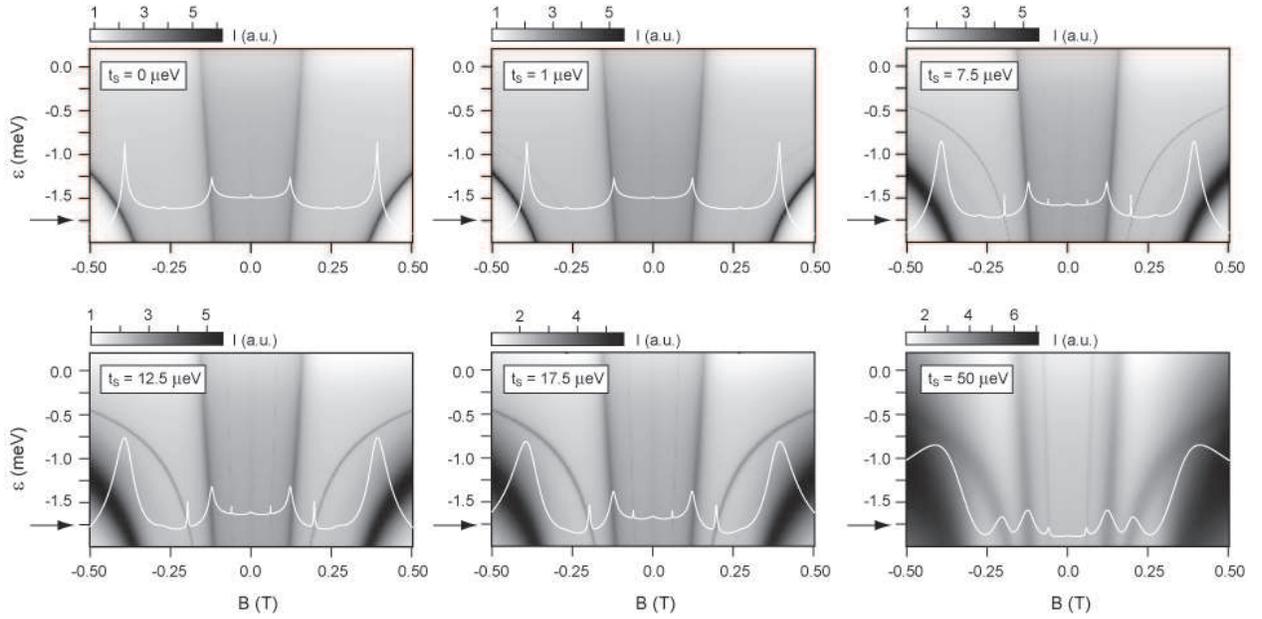}
\begin{doublespace}
\caption{\label{figS3}As Fig.3(c)(right) in the main article for different spin-flip tunnel couplings.}
\end{doublespace}
\end{figure}

Figure~S\ref{figS3} shows the current for different spin-flip
tunnel couplings between the two quantum dots. As explained in the
main article with increasing coupling additional high current
curves appear. Further, the width of the curves increases with the
spin-flip coupling. This is attributed to coherent transitions
between the three-spin states, which are induced solely when the
spin-flip coupling is nonzero. These transitions act in addition
to incoherent transitions induced by spin-flips because of the
coupling to the heat baths. Both effects define a limit on the
magnitude of the spin-flip coupling observed in the experiments.
For a too weak coupling no additional curves would be observed,
whereas for a too strong coupling the curves would be washed out
as shown in Fig.~S\ref{figS3}. The theoretical model indicates that
the spin-flip coupling should be on the order of $t_{s}\sim 10$
$\mu$eV and the best fit to the experimental data is obtained for
$t_{s}\approx 7.5$ $\mu$eV.

\end{doublespace}


\end{document}